\documentclass[prx,twocolumn,showpacs,amsmath,amssymb,superscriptaddress]{revtex4-2}
\usepackage{graphicx}
\usepackage{dcolumn}
\usepackage{bm}
\usepackage{hyperref}
\hypersetup{colorlinks,citecolor=blue, filecolor=blue, linkcolor=blue , urlcolor=blue}
\usepackage{ulem}

\begin{document}

\title{Deconstruction of the anisotropic magnetic interactions from spin-entangled optical excitations in van der Waals antiferromagnets}

\author{Dipankar Jana}
    \email{jana.d02@nus.edu.sg}
    \affiliation{LNCMI-EMFL, CNRS UPR3228, Univ. Grenoble Alpes, Univ. Toulouse, Univ. Toulouse 3, INSA-T, Grenoble and Toulouse, France}
    \affiliation{Institute for Functional Intelligent Materials, National University of Singapore, 117544, Singapore}

\author{Swagata Acharya}
    \email{swagata.acharya@nrel.gov}
    \affiliation{National Renewable Energy Laboratory, Golden, CO, 80401 USA}

\author{Milan~Orlita}
    \affiliation{LNCMI-EMFL, CNRS UPR3228, Univ. Grenoble Alpes, Univ. Toulouse, Univ. Toulouse 3, INSA-T, Grenoble and Toulouse, France}
    \affiliation{Institute of Physics, Charles University, Ke Karlovu 5, Prague, 121 16, Czech Republic}

\author{Clement~Faugeras}
    \affiliation{LNCMI-EMFL, CNRS UPR3228, Univ. Grenoble Alpes, Univ. Toulouse, Univ. Toulouse 3, INSA-T, Grenoble and Toulouse, France}
    
\author{Dimitar~Pashov}
\affiliation{King’s College London, Theory and Simulation of Condensed Matter, The Strand, WC2R 2LS London, UK}   

\author{Mark van Schilfgaarde}
    \affiliation{National Renewable Energy Laboratory, Golden, CO, 80401 USA}

\author{Marek~Potemski}
    \email{marek.potemski@lncmi.cnrs.fr}
    \affiliation{LNCMI-EMFL, CNRS UPR3228, Univ. Grenoble Alpes, Univ. Toulouse, Univ. Toulouse 3, INSA-T, Grenoble and Toulouse, France}
     \affiliation{CENTERA, CEZAMAT, Warsaw University of Technology, 02-822 Warsaw, Poland} 
    \affiliation{Institute of High Pressure Physics, PAS, 01-142 Warsaw, Poland} 

\author{Maciej~Koperski}
    \email{msemaci@nus.edu.sg}
    \affiliation{Institute for Functional Intelligent Materials, National University of Singapore, 117544, Singapore}
    \affiliation{Department of Materials Science and Engineering, National University of Singapore, 117575, Singapore}

\begin{abstract}

Magneto-optical excitations in antiferromagnetic \textit{d} systems can originate from a multiplicity of light-spin and spin-spin interactions, as the light and spin degrees of freedom can be entangled. This is exemplified in van der Waals systems with attendant strong anisotropy between in-plane and out-of-plane directions, such as MnPS$_3$ and NiPS$_3$ films studied here. The rich interplay between the magnetic ordering and sub-bandgap optical transitions poses a challenge to resolve the mechanisms driving spin-entangled optical transitions, as well as the single-particle bandgap itself. Here we employ a high-fidelity \textit{ab initio} theory to find a realistic estimation of the bandgap by elucidating the atom- and orbital-resolved contributions to the fundamental sub-bands. We further demonstrate that the spin-entangled excitations, observable as photoluminescence and absorption resonances, originate from an on-site spin-flip transition confined to a magnetic atom (Mn or Ni). The evolution of the spin-flip transition in a magnetic field was used to deduce the effective exchange coupling and
anisotropy constants. 
\end{abstract}

\maketitle

\section{Introduction}

Understanding the impact of the macroscopic spin arrangement in magnetic materials on possible optical transitions constitutes a fundamental requirement for devising protocols for probing and manipulating magnetic states with light \cite{Disa2023, Ilyas2024, Khusyainov2023t, Gao2023, Afanasiev2021}. Spin polarization from partially filled \textit{d} orbitals in transition metal insulators strongly influences the one-particle properties participating in the optical transitions~\cite{Tanabe1954} as well as the bandgap. The interplay between magnetism and one-particle properties is also the
primary mechanism responsible for collective magnetic order.  These orbitals are often localized, leading to flat band dispersions, heavy electrons, and large spin scattering, which favors the emergence of correlated electronic properties \cite{CrX3_correlated, CrX3_spin_textures}. Consequently, the interplay between the magnetic order and the optical excitations may be complex and strongly dependent on the individual characteristics of a particular system.

Here, we present evidence of coupling between the magnetic order and the optical transitions in two representative van der Waals antiferromagnets, MnPS$_3$ and NiPS$_3$. These two systems exhibit out-of-plane and in-plane easy spin axes, respectively. The effective exchange interaction, which determines the magnetic order, exhibits antiferromagnetic character in both systems. This spin arrangement in the lattice directly contributes to the magnon gap excitation revealed by Raman scattering and absorption processes~\cite{kobets2009, Dipankar2023} and couples with radiative resonances observable in photoluminescence and absorption spectra~\cite{Gnatchenko2011, Afanasiev2021, Wang2021, Dipankar2023}. The magnon modes and the radiative resonance exhibit Zeeman splitting~\cite{kobets2009, Gnatchenko2011, Dipankar2023}, which depends on the canting of the field relative to the easy axis. The magnetic field evolution of the radiative transition provides sufficient information to infer key parameters controlling magnetic interactions, such as \textit{g}-factors, spin-flip critical magnetic fields, effective exchange coupling, and anisotropy, which determine the magnetic order in the material. These data are in agreement with previous results of neutron scattering characterization \cite{Wildes1998, Wildes2022} and extend the capabilities of probing antiferromagnetism to thin microscale samples with significantly smaller magnetic domain features through optical microscopy methods. 

Using a self-consistent form of \textit{ab initio} many-body perturbation theory (MBPT) and locally exact dynamical mean field theory (DMFT), we compute the spin-allowed and spin-flip transitions, respectively. We unambiguously show that the spectrally narrow optical excitations arise from spin-flip transitions occurring at the magnetic transition metal sites (Mn or Ni) and, hence, the transitions have a one-to-one analogy with the spin-forbidden transitions in the Tanabe-Sugano diagram.  However, our calculations also explore transitions in MnPS$_3$ and NiPS$_3$ over a wide energy window, including spin-flip, non-spin-flip on-site, dipolar inter-site, and charge transfer transitions. Observed optical properties encompass broadband low-energy transitions involving spin-allowed intra-\textit{d}-shell transitions and higher-energy charge transfer excitons at the ultraviolet window for MnPS$_3$ and visible window for NiPS$_3$. 

\begin{figure*}[t]
	\includegraphics[width=16.8cm]{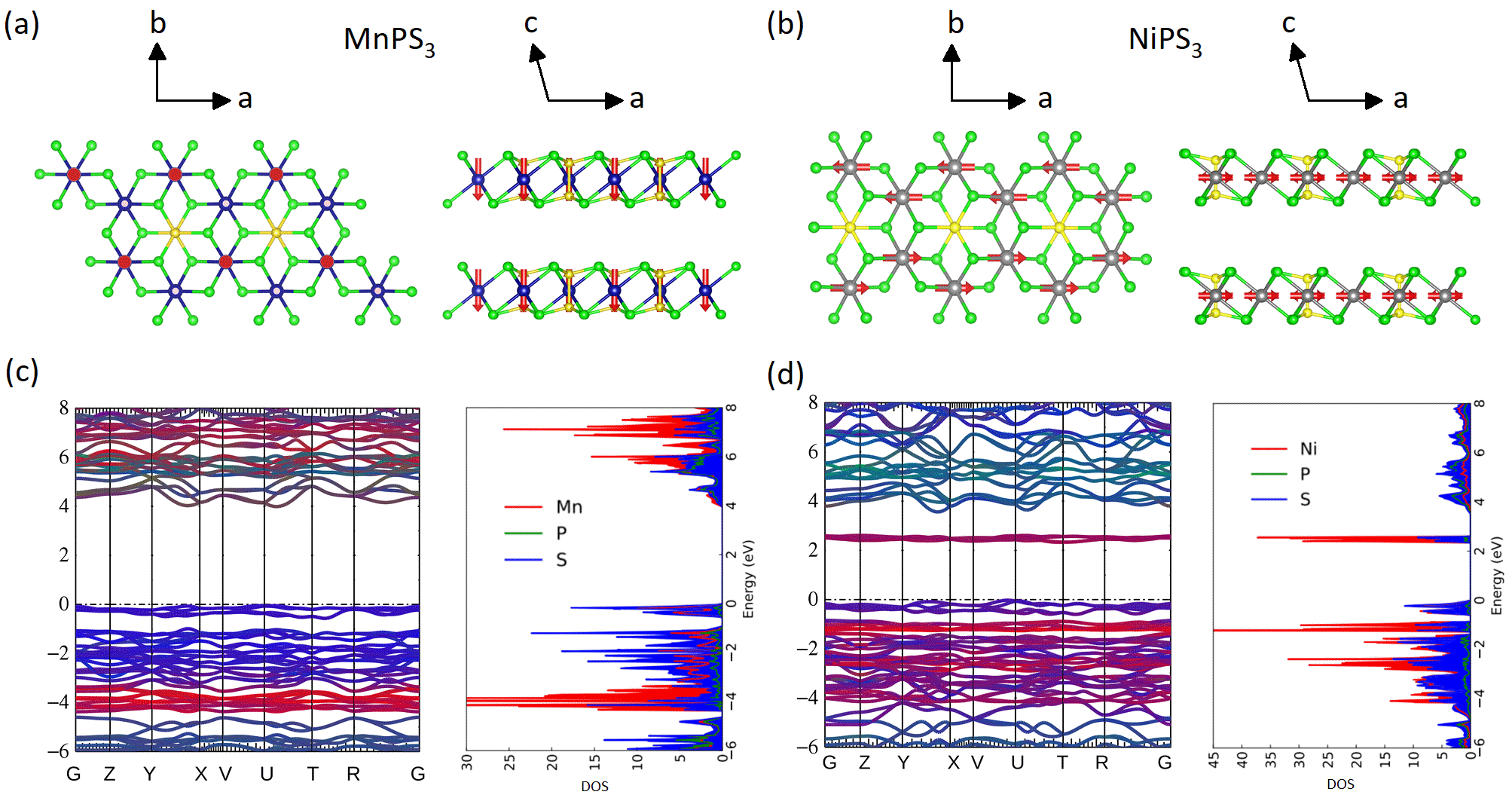}
	\caption{Schematic crystallographic and magnetic structure of (a) MnPS$_3$ and  (b) NiPS$_3$, highlighting the spin orientations in the antiferromagnetic phase. Blue/gray/green/yellow spheres represent manganese/nickel/sulfur/phosphorus atoms. These figures are created using the VESTA software package~\cite{Momma2011}. The electronic band structure and atom-projected density of states for the bulk (c) MnPS$_3$ and (d) NiPS$_3$ calculated with the and QS${G\hat{W}}$ method. The Fermi level is set to zero.}
	\label{fig:fig1}
\end{figure*}

These findings contribute to a systematic understanding of the sub-bandgap optical transitions in the MPS$_3$ family of materials. Their magneto-optical characterization establishes protocols for all-optical probing of the emergent unique spin configuration in antiferromagnets.

\section{Results}
\subsection{Electronic band structure}

First, we compute the electronic structures of MnPS$_3$ and NiPS$_3$ bulk crystals within the self-consistent \textit{ab initio} MBPT, quasi-particle self-consistent \textit{GW} (QS$GW$) \cite{qsgw} and QS$G\hat{W}$ \cite{Cunningham2023} approaches. QS$G\hat{W}$ is a self-consistent extension of QS$GW$ \cite{qsgw,questaal_paper} where electronic eigenfunctions are computed in the presence of the screened Coulomb correlations corrected by the excitonic vertex. Self-consistency is imposed for both the self-energy $\Sigma$ and the charge density. The latter is usually neglected in \textit{GW}, however, it has been demonstrated to modify the electronic structure for a certain class of materials systems  \cite{acharya2021importance,tise2}. Self-consistency in $\Sigma$ ensures that feedback between the one-particle spectrum and the Coulomb interactions is accounted for~\cite{Vidal10}.  This is particularly important when magnetic degrees of freedom are involved, as there is an additional coupling between spin and $\Sigma$.  $G$, $\Sigma$, and $\hat{W}$ are updated iteratively until all of them converge.  Our results are thus parameter-free and have no starting point bias. The monoclinic crystal structures of bulk MnPS$_3$ and NiPS$_3$, characterized by C2/m space symmetry, are presented schematically in \textbf{Fig.~1(a,b)}, together with the antiferromagnetic spin alignment in the magnetic phase. Mn/Ni atoms are arranged in hexagonal patterns within planes coupled via van der Waals forces. The easy spin axis is out-of-plane for MnPS$_3$ and in-plane for NiPS$_3$. The corresponding $QSG\hat{W}$ band structures and atom-projected density of states (DOS) are shown in \textbf{Fig.~1(c,d)}. The bulk electronic bandgap ($E_g$) for MnPS$_3$ was found to be 1.1\,eV in local-density approximation (LDA), 4.2\,eV in QS$GW$ and 3.9\,eV in QS$G\hat{W}$.  For NiPS$_{3}$ they are respectively 0.8\,eV (LDA), 2.5\,eV (QS$GW$) and 2.2\,eV (QS$G\hat{W}$).  The large enhancement in bandgaps within QS$GW$ (compared to LDA) and a subsequent 10-15\% reduction in QS$G\hat{W}$ when excitonic correlations screen $W$ follows a typical pattern for transition metal magnetic insulators~\cite{Cunningham2023,acharya2021electronic,bianchi2023paramagnetic,watson2024giant}. The atom-projected DOS shows that both materials have charge-transfer electronic structures with S-\textit{p} states dominating the valence band top. Ni and Mn \textit{d} states align differently to the S-\textit{p} state because of two factors. First, an atomic Ni \textit{d} state is deeper than the Mn counterpart because it senses a stronger nuclear attraction, making the average of the $(d^{\uparrow},d^{\downarrow})$ band center deeper (in the material, electrostatic shifts from charge transfer reduce the free-atomic Ni-Mn difference).  On the other hand, the $(d^{\uparrow},d^{\downarrow})$ spin splitting is smaller, because Ni has a much smaller exchange splitting.  The \textit{d} band center is spin split by approximately ${\pm}I{\cdot}M$ where $M$ is the local moment and $I$ the Stoner parameter, known to be $~\sim$1\,eV in the 3\textit{d} metals.  QS$G\hat{W}$ calculations yield local moments of 1.45\,$\mu_B$ and 4.7\,$\mu_B$ on Ni and Mn, respectively. Thus, we expect majority-minority spin splitting of \textit{d} states to be $\sim$2.9\,eV, and $\sim$9.4\,eV in Ni and Mn, respectively.  The spin-averaged positions and the spin splitting approximately follow the expected trends, although crystal-field effects complicate the true situation. The center of mass of Mn \textit{d} states is located 4 eV below the valence band edge $E_{v}$ while for Ni it is 1 eV below, whereas the unoccupied \textit{d} states are at $E_{v}{+}7$\,eV for MnPS$_3$ and $E_{v}{+}2$\,eV for NiPS$_3$.
The suppression of magnetic moments in these two systems is driven by the $pd$ charge transfer from ligand to metal, which screens the ionic moment. The efficiency of this mechanism can be linked to the symmetry of the crystal lattice. A similar situation is observed in NiO, where the magnetic moment is found to be 1.7~$\mu_B$~\cite{acharyaTheoryColorsStrongly2023}, smaller than the ionic moment of 2 $\mu_B$. The discrepancy in NiPS$_{3}$ is even larger because of lowered crystal-field symmetry. NiPS$_{3}$ belongs to a monoclinic space group with crystal-field symmetry significantly lower than the cubic crystal-field of NiO. In NiO, the cubic crystal-field ensures that the spin configuration of Ni$^{2+}$ is t$_{2g}$$^{6}$e$_{g}$$^{2}$. The monoclinic distortion in NiPS$_{3}$ implies that all $d$ orbitals are partially occupied and the net moment is smaller. A Mulliken decomposition of the orbital occupancies in the two spin channels suggests that the m$_{l}$=-2, -1, 0, 1, 2 states contribute respectively 0.24, 0.41, 0.15, 0.41, 0.24 $\mu_{B}$ to the net moment of 1.45 $\mu_{B}$, in contrast to the expected contributions of 1 $\mu_{B}$ each from the m$_{l}$=0 and m$_{l}$=2 states. We also find that of the 1.45 $\mu_{B}$, 1.33 $\mu_{B}$ is concentrated inside the augmented sphere around the Ni atom, while the remaining 0.12 $\mu_{B}$ resides on the metal-ligand bond. For MnPS$_{3}$ the distribution of atomic and bonded parts of the magnetic moment is 4.03 and 0.67 $\mu_{B}$, respectively.

Rich excitonic spectra in both systems are computed with the two different approaches noted above: in MBPT within $\mathrm{QS}G\hat{W}$ framework, and by exact computation of higher order charge-charge correlators within DMFT, using an exact-diagonalization (ED) impurity solver~\cite{acharyaTheoryColorsStrongly2023}. $\mathrm{QS}G\hat{W}$ is limited to spin-allowed excitonic transitions, but the nonperturbative, locally exact impurity vertex in ED-DMFT produces the spin-flip atomic multiplet transitions missing from MBPT. The combined approach has been shown to produce the mid-gap (in the visible range) spin-allowed and spin-flip transitions that lead to the green and pink colors of MnF$_{2}$ and NiO, respectively~\cite{acharyaTheoryColorsStrongly2023}. The remarkable potential of the combined approach lies in its ability to unambiguously compute and assign spin-allowed and spin-forbidden (flip) characters to the excitonic states. The computed exciton spectra for MnPS$_3$ and NiPS$_3$ are shown in \textbf{Fig.~2}. For MnPS$_{3}$, $\mathrm{QS}G\hat{W}$ produces spin-allowed excitonic transitions $E_{ex}$ at 3.2\,eV and above. This is explained in terms of the Mn $d^{5}$ occupation, with the majority sector fully occupied and the minority sector fully unoccupied.  In such a configuration, any atomic on-site transition leads to a spin-flip process, which is not captured by $\mathrm{QS}G\hat{W}$ as it only produces excitonic transitions which are either $pd$ or intersite $dd$ in character. The situation is different in NiPS$_{3}$. Ni is characterized by a $d^{8}$ state, allowing atomic intra-site transitions between t$_{2g}$ and e$_{g}$ states that do not require a spin flip. Consequently, $\mathrm{QS}G\hat{W}$ predicts a 1\,eV transition in NiPS$_{3}$ which is a spin-allowed atomic transition on the Ni atom. Additionally, ED-DMFT captures transitions involving spin flips at the energy of 2.64\,eV in MnPS$_{3}$ and 1.47\,eV in NiPS$_{3}$.  The 2.64\,eV transition excites the ground state atomic spin configuration from $5/2$ to $3/2$ in Mn, and the 1.47\,eV transition changes the spin state from $1$ to $0$ in Ni, both involving strictly spin-flip processes on the atomic sites. By employing complementary theoretical methodology, we were able to distinguish between the extended excitonic states and atom-local Frenkel states for both the systems (see the S.~I.~Appendix for the visualization of the spin-allowed extended states in Mn and spin-forbidden atom-local exciton wave functions).

\begin{figure*}[t]
	\includegraphics[width=17cm]{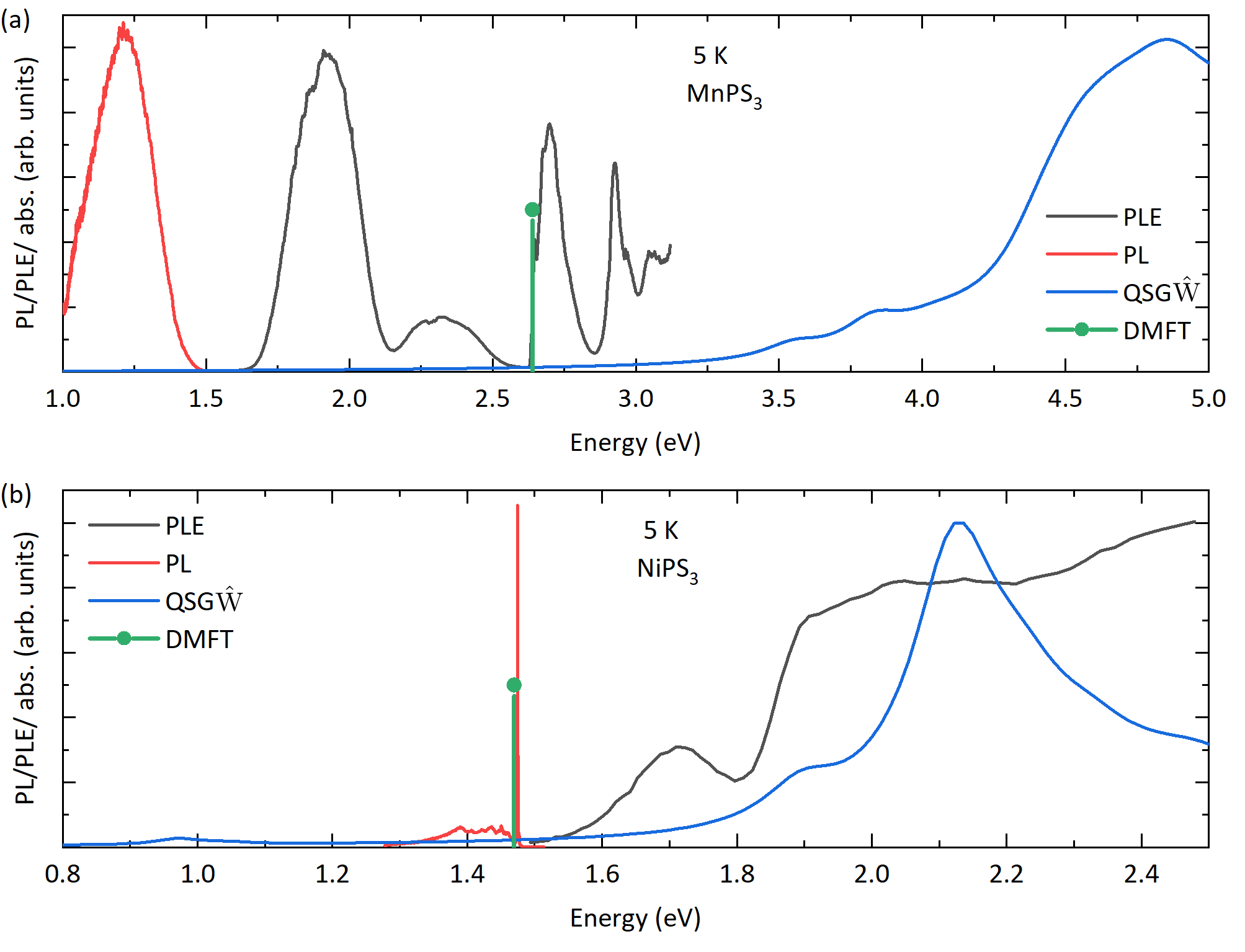}
	\caption{Low-temperature (5~K) PL and the PL excitation spectra (PLE) corresponding to the dominant PL feature of (a) MnPS$_3$ and (b) NiPS$_3$. The computed optical absorption spectra by $\mathrm{QSG\hat{W}}$ and DMFT methods are also shown in the corresponding plot.}
	\label{fig:fig2}
\end{figure*}

\subsection{Broadband optical response}
Within our experimental approaches, we probe the same transitions via photoluminescence (PL) and photoluminescence
excitation (PLE) spectroscopy. Such spectra for bulk MnPS$_3$ and NiPS$_3$ at 5 K (below the critical N\'{e}el
temperature), presented in \textbf{Fig.~2}, are indicative of the emission and quasi-absorption characteristics of the
materials. Due to the limited experimentally investigated spectral range, we could not probe the bandgap and the
spin-allowed excitonic states of MnPS$_3$. Instead, we observe several broad emission and absorption bands below the
predicted spin-allowed excitonic transition, indicating the coexistence of optical transitions of different microscopic
origins. The energies of the pronounced excitation transition at 1.9,~2.3, and 2.6\,eV have been reported before and
argued to have explanations in the ligand field theory description in the form of the Tanabe-Sugano diagram for an
Mn$^{2+}$ ion~\cite{Grasso1991}. They correspond to the on-site excitations in the Mn ion, which cause reconfiguration of the valence \textit{d}-shell electronic spins from half-occupied orbitals to fully-occupied singlet states, effectively involving a
spin-flip in this process~\cite{Tanabe1954}. The modification of the electronic spin configuration significantly impacts
the interatomic bonding strength, leading to the local distortion of the crystal structure. The resulting modification
of the structural energy favors the coupling of the electronic transition to phonons, leading to the emergence of
spectrally broad absorption bands. The broad emission at 1.2\,eV results from multiple-phonon-assisted Stokes-shift
emission, corresponding to the 1.9\,eV absorption band, and can be described using the configuration coordinate diagram~\cite{Dexter1955}. The PLE spectra of NiPS$_3$ exhibit broadband transitions in the vicinity of the single-particle band gap, suggesting a contribution from spatially delocalized near-band-edge spin-allowed exciton as
predicted by the $\mathrm{QS}G\hat{W}$. Since an excitonic state at 1.8\,eV is only moderately bound, it is natural
that this state has an extended character. The broad PL at 1.4\,eV is the Stokes-shifted emission from the 1.7\,eV
spin-allowed exciton state. The observed optical transitions in PL/PLE are net products of a strong coupling between the electronic degrees of freedom and the bosonic modes, including phonons and magnons. This leads to a significant broadening and Stokes shift of the sub-band gap transitions, which often masks the true band edge of the system in PL/PLE data. Our theoretically computed single particle bandgap of NiPS$_3$ is additionally in agreement with the electronic interband resonances deduced from the electron scattering and optical conductivity spectra~\cite{Klaproth2023, Kim2018}.


\begin{figure*}[bt]
	\includegraphics[width=17cm]{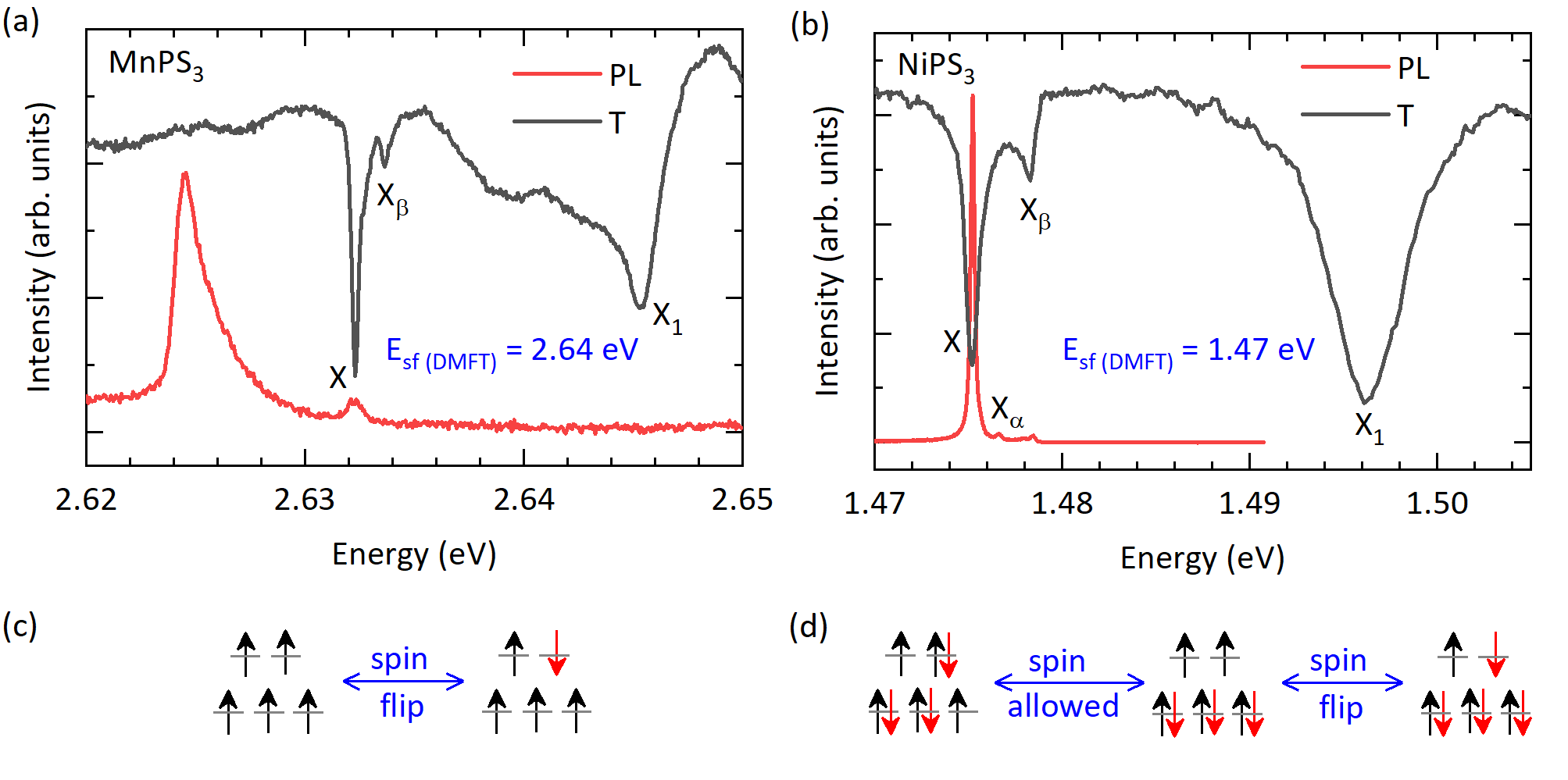}
	\caption{Low temperature (5~K) PL and transmission (T) spectra of (a) MnPS$_3$ and (b) NiPS$_3$ shown in a narrow energy range near the spin-flip resonance, labeled as X. The high-energy satellite peaks X$_\alpha$ and X$_\beta$ correspond to the phonon replica of X-transition, while X$_1$ peak corresponds to the exciton-magnon continuum coupled state. The schematic spin arrangements of the ground state, onsite spin-allowed, and spin-flip excited states of (c) MnPS$_3$ and (d) NiPS$_3$. No onsite spin-allowed transition is feasible in MnPS$_3$ due to half-filled spin arrangements.}
	\label{fig:fig3}
\end{figure*}

In addition to these broad transitions, the PL spectrum of NiPS$_3$ is dominated by a spectrally narrow resonance. A weak signature of a similar narrow resonance appears in the PL spectra of MnPS$_3$ when excited with a UV laser (385\,nm). The narrow resonance in the PL spectra is further supported by transmission spectra shown in \textbf{Fig.~3(a,b)}. The transition labeled as X is reported to be spin-entangled, i.e., its oscillator strength, peak energy, and polarization are coupled to the magnetic order \cite{Kang2020, Dipankar2023}. This coupling offers insight into the evolution of the magnetic state with external parameters such as temperature and magnetic field. In the framework of the Tanabe-Sugano diagram \cite{Tanabe1954}, this transition can be considered qualitatively as the onsite transition of the metal ion when a spin-flip occurs in one of the half-occupied orbitals while preserving the number of half-occupied and fully occupied states as illustrated schematically in \textbf{Fig.~3(c,d)}. As shown before, our ED-DMFT approach finds
transitions at 2.64\,eV for MnPS$_3$ and 
at 1.47\,eV for NiPS$_3$, as indicated in \textbf{Fig~2(a,b)}, confirming the origin of X-feature as a spin-flip d-d excitation within a half-occupied orbital of the metal ion. 
The origin of this transition of NiPS$_3$ is debated and interpreted in various ways \cite{Kang2020, Hwangbo2021, Klaproth2023, Kim2023, song2024manipulation, He2024, Hamad2024}. Such ultra-narrow sub-bandgap transitions have been observed in various materials~\cite{Gnatchenko2011, Kang2020, Son2022, Occhialini2024, van1967optical}. Our theory demonstrates that they can be broadly classified as onsite spin-flip d-d transitions. In Section~4 of the SM~\cite{SuppInfo}, we have discussed the characteristics of the 1.47~eV spin-flip transition in NiPS$_3$ in the context of distinct theoretical models and their applicability to the experimental data presented herein and reported in literature.

\begin{figure*}[ht]
	\includegraphics[width=17cm]{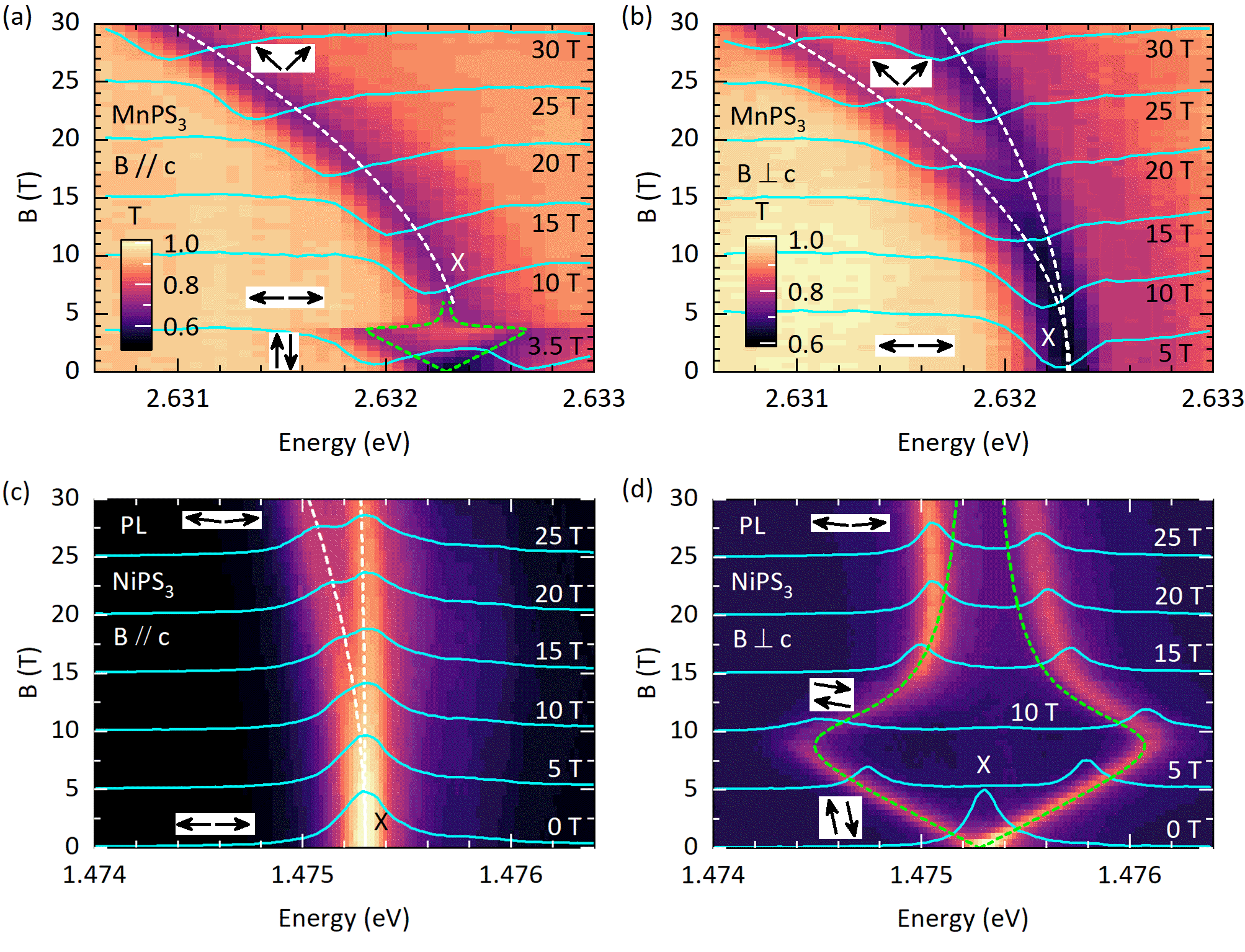}
	\caption{False color map of transmission of bulk MnPS$_3$ as a function of the (a) out-of-plane ($B \parallel c$~-axis) and (b) in-plane ($B \perp c$~-axis) magnetic field. False color map of PL of bulk NiPS$_3$ as a function of the (c) out-of-plane ($B \parallel c$~-axis) and (d) in-plane ($B \perp c$~-axis) magnetic field. Representative spectra at 5~T intervals are shown in the plot. The Green dashed line corresponds to the simulation of the splitting of the X-transition with low field approximation~\cite{Dipankar2023} ($J \gg g\mu _BB$) while the white dashed line corresponds to the simulation in the high field limit following Eq.~\ref{eq:1}. Black arrows show the schematic spin alignment of the two spin sub-lattices for the external magnetic field applied along the vertical axis (B(T) axis).}
	\label{fig:fig4}
\end{figure*}

\subsection{The deconstruction of the magnetic interaction parameters}
The external magnetic field compensates for the antiferromagnetic exchange coupling (J) leading to modification in the spin orientation \cite{Guo2022}. The qualitative evolution of the spin-flip transition depends on the relative alignment of the direction of a magnetic field and the easy spin axis characterizing the magnetic order \cite{Dipankar2023, Wang2024}. This is illustrated in \textbf{Fig.~4} by the magneto-transmission and magneto-PL spectra measurement in two configurations: magnetic field aligned in the same plane as the easy spin axes (B $\parallel$ c for MnPS$_3$ and  B $\perp$ c for NiPS$_3$) and magnetic field aligned perpendicular to the easy spin axis (B $\perp$ c for MnPS$_3$ and  B $\parallel$ c for NiPS$_3$). As the spin-flip resonance is spin-entangled, it exhibits a Zeeman effect according to the formula~\cite{Wang2024}:

\begin{equation} \label{eq:1}
E_X (B)= E_X (B=0~T) - g\mu_BB\cos\Psi(B)  
\end{equation}

where E$_X(B = 0~T)$ is the zero-field energy of the spin-flip exciton, $g$ is the Land\'{e} \textit{g}-factor, $\mu_B$ is Bohr magneton, and $\Psi$(B) is the angle between the direction of a magnetic field and spin axis of the sublattices. The spin selection for the spin-flip process is considered to be $\Delta S=1$. In the first scenario (presented in \textbf{Fig.~4(a,d)}), when the magnetic field is applied along the easy spin axis ($\Psi(B)=0^{\circ}, 180^{\circ}$ for the two spin sub-lattices), the exciton splits into two components, varying linearly with the external magnetic field. Under the conditions that the effective exchange coupling constant (J) is significantly larger than the anisotropy parameter (D), i.e., J $\gg$ D, such evolution continues until the spin sublattices find another equilibrium position, aligning perpendicular to the field direction ($\Psi(B)=90^{\circ}$ for both spin sub-lattices) but maintaining the antiferromagnetic ordering as schematically shown in the insets of \textbf{Fig.~4(a,d)} \cite{Nagamiya1955}. The spin rotation is directly reflected in the rotation of the linear polarization axis as shown in \textbf{Fig.~S2}~ of SM~\cite{SuppInfo}. This spin reorientation causes the Zeeman splitting to collapse within the field interval designated as a spin-flop field (B$_{sf}$) \cite{Gnatchenko2011, Dipankar2023}. We were not able to align the easy spin axis exactly along the field, resulting in a gradual convergence of the split components of X-transition. Simulation for the magnetic field dependence, considering a finite angle between the magnetic field and the easy spin axis~\cite{Nagamiya1955} (3$^{\circ}$ for MnPS$_3$ and 15$^{\circ}$ for NiPS$_3$) are shown by a green dashed line in \textbf{Fig.~4(a,d)}. A spin-flop field of 3.8~T with a \textit{g}-factor of 1.93 for MnPS$_3$ and of 10.5~T with a g-factor of 2.00 for NiPS$_3$ were estimated from the simulation. 

Above the spin-flop field, the X-transition redshifts without splitting. A quadratic redshift of the spin-flip resonance as a function of the magnetic field was reported recently for NiPS$_3$ and justified by the presence of a dark exciton state above the X-transition \cite{Wang2024}. However, this quadratic shift indicates that the external magnetic field starts canting both the sublattice spins towards the field direction. Within the J $\gg$ D assumption, $\Psi$ (B) can be expressed as~\cite{cho2023}:

\begin{equation} \label{eq:2}
\cos\Psi (B) = \frac{g\mu_BB}{2S_gJ}
\end{equation}

where $S_g$ corresponds to the total spin of the ground state (S$_g$=5/2 for MnPS$_3$ and S$_g$=1 for NiPS$_3$). The simulated quadratic magnetic field dependence is shown in \textbf{Fig.~4(a,d)} and in \textbf{Fig.~S3} by the dashed line. This analysis enables us to determine the effective exchange constant $J$ to be 1.6~meV for MnPS$_3$. The spin of the sublattices will become parallel to the field direction, marking the saturation of magnetization ($ \cos \Psi (B)=1$) when $g\mu_BB_c$=2S$_g$J, where B$_c$ corresponds to the critical saturation field. Using the estimated value of $J$, $B_c$ is predicted to be 71~T for MnPS$_3$. On the other hand, $B_{sf}$ is determined by both the effective exchange coupling constant and the anisotropy parameter according to the relationship arising from the direct diagonalization of the spin Hamiltonian as reported previously~\cite{Rezende2019}:
\begin{equation} \label{eq:3}
D=(g\mu_BB_{sf})^2/8S_g^2J
\end{equation}

the anisotropy parameter $D_{z}$ for MnPS$_3$ was estimated to be 0.002~meV, which is consistent with our previous assumption that the anisotropy energy is much smaller than the 
 effective exchange energy. For NiPS$_3$, no significant quadratic shift was observed in the high-field limit in this configuration. This is primarily obscured by the gradual convergence of the split components and also suggests a much larger effective exchange constant when compared to MnPS$_3$.

In order to establish the magnetic interaction parameters for NiPS$_3$, we analyze the magneto-optical spectra in the second configuration. The dependence of X-transition on the magnetic field applied perpendicular to the easy spin axis (B $\perp$ c for MnPS$_3$ and  B $\parallel$ c for NiPS$_3$) is similar to the field dependence in the canted phase after the spin-flop field (B$_{sf}$). The evolution of the X-transition in this configuration of the applied magnetic field is shown in \textbf{Fig.~\ref{fig:fig4}(b,c)} for MnPS$_3$ and NiPS$_3$, respectively. We see the splitting of this transition into two components, one of which shows an identical quadratic redshift to that observed for the case of B applied parallel to the easy spin axis (see also \textbf{Fig.~S3} of SM). Similar splitting of the ground state doublet for the magnetic field applied perpendicular to the spin direction was reported before~\cite{joy1992} for NiPS$_3$ where the energy of one component changed quadratically while the other component remained independent of the magnetic field. However, our high-field data revealed that the high-energy component was also redshifted quadratically with a relatively lower rate. In the case of NiPS$_3$, the quadratic redshift of the low energy split component was reproduced well by our model as demonstrated in \textbf{Fig.~4(c)} by the dashed line (see also \textbf{Fig.~S4} of SM). In the case of NiPS$_3$, we obtained the value of the effective exchange coupling constant $J = 22.5$~meV, the critical saturation field $B_c = 390$~T, and the in-plane anisotropy constant $D_x = 0.008$~meV. Notably, NiPS$_3$ is a bi-axial antiferromagnet characterized by a small in-plane anisotropy and relatively strong out-of-plane anisotropy \cite{Wildes2022, Dipankar2023}. The anisotropy parameter estimated here corresponds to the low-energy in-plane component. The parameters estimated from the magnetic field dependence of the X-transition are tabulated in \textbf{Tab.~\ref{Table:1}}. The effective exchange parameter (J) estimated in this analysis is the sum of all pairwise exchange interactions to the third nearest neighbor within the layer. The interlayer exchange interaction parameter is reported to be small and thus, neglected~\cite{yagotintsev2012interlayer, Wildes2022}. Slight thickness-dependent modifications in N\'eel temperature, which is related to the effective exchange interaction, were observed previously~\cite{lim2021thickness, kim2019suppression}. Thus, we do not expect any strong thickness dependence of the interaction parameters. Nevertheless, the significant difference in the interaction parameters and the values of the critical magnetic fields highlight the intrinsic anisotropies in van der Waals antiferromagnets when the easy spin axis is oriented along the direction of weak van der Waals forces (as in the case of MnPS$_3$) or along the direction of strong covalent bonding forces (as in the case of NiPS$_3$).

\begin{table}[]
\begin{tabular}{|c|c|c|c|c|c|}
\hline
      & \textit{g}-factor & B$_{sf}$ (T) & J (meV)    & B$_c$ (T)  & D (meV)     \\ \hline
MnPS$_3$ & 1.93        & 3.8 & 1.6~~~~~  & 71  & 0.002 (z)~ \\
 & & & ~3.2~\cite{Wildes1998} & & 0.009~\cite{Wildes1998} \\
 & & & 2.3~$^*$~~ & & \\
\hline
NiPS$_3$ & 2        & 10.5  & 22.5~~~~~~ & 390 & 0.008 (x)~ \\ 
 & & & 38.7~\cite{Wildes2022} & & 0.010~\cite{Wildes2022} \\
 & & & 20.1~$^*$~~~ & & \\
\hline
\end{tabular}
\caption{The magnetic interaction parameters for MnPS$_3$ and NiPS$_3$ bulk films estimated from the magnetic field dependence of the spin-flip resonance excitation. The anisotropy constant (D) corresponds to the low-energy in-plane anisotropy (D$_x$) for the biaxial antiferromagnet NiPS$_3$ and out-of-plane anisotropy (D$_z$) for the uniaxial antiferromagnet MnPS$_3$. For comparison, values of the $J$ and $D$ parameters from previous studies (Ref.~\cite{Wildes1998} and Ref.~\cite{Wildes2022}), which account for exchange interactions up to the third nearest neighbor, are also provided. It is calculated using the formula: $J=z_1J_1+z_2J_2+z_3J_3$, where z$_i$ and J$_i$ represent the number i$^{th}$ nearest neighbors and the corresponding exchange interaction, respectively. Values marked with $^*$ denote effective exchange parameters estimated from the N\'eel temperature using the relation: $J=3 k_B T_N /[S (S+1)]$ as described in Ref.\cite{Basnet2022}.}
\label{Table:1}
\end{table}

\section{Conclusions}
In conclusion, the optical characterization of MnPS$_3$ and NiPS$_3$ bulk films revealed rich spectroscopic characteristics encompassing recombination from electronic states characterized by varied levels of localization and diverse spin configurations. We demonstrated that the magnetic field evolution of the spin-entangled on-site $dd$ spin-flip excitation contains information on the magnetic interactions, which determine the antiferromagnetic order in van der Waals systems. The entanglement between optical excitation and magnetism, impacted by the intrinsic anisotropies of the two-dimensional systems, can lead to multifaceted functionalities related to ultrafast probing and manipulation of the magnetic states with light, or conversely reflecting the magnetic state of the sample in the polarization properties of emitted or absorbed photons.

\section{Experimental Section and Computational Methods}
\textit{Samples:} Commercially available samples from HQ graphene are used in this study. A bulk MnPS$_3$ sample (3~mm~by~3~mm) is used for the reflectivity measurements, while thin exfoliated flakes (100~\textmu m~by~100~\textmu m) of NiPS$_3$ are used for PL measurements.

\textit{Optical measurements:} For transmission measurements, light from a quartz tungsten halogen lamp is coupled to a free beam probe by optical fiber and focused on the sample's bottom surface with a microscope objective of NA=0.83. The reflected (double-transmitted) signal is collected by the same objective, dispersed by a 0.75 m monochromator (2000~l/mm grating), and detected by a charge-coupled device camera, which is cooled at 120 K. For PL measurement, an identical setup is used except for an excitation from a 515~nm continuous laser. Appropriate filters are used before the excitation to clean the laser line/ to block the second-order effect and before the collection to block the laser from reaching the monochromator. The measurements are performed at liquid helium temperature, where the sample was placed inside a liquid resistive magnet which can reach up to 30~T.

\textit{Computational methodology:} For both bulk NiPS$_{3}$ and MnPS$_{3}$ in the AFM phase, a 20-atom unit cell is used. The single particle calculations (LDA, and energy band calculations with the static quasiparticlized $\mathrm{QSGW}$ and $\mathrm{QSG\hat{W}}$ $\Sigma$(k)) are performed on a 6$\times$4$\times$4 k-mesh while the relatively smooth dynamical self-energy $\Sigma(\omega)$ is constructed using a 3$\times$2$\times$2 k-mesh. However, we also observe that the difference in electronic band gaps between 6$\times$4$\times$4 k-mesh and 3$\times$2$\times$2 k-mesh is only about $40$ meV, which is a small correction considering its band gap value of $\sim$2.2 eV and $\sim$4 eV, respectively.  The $\mathrm{QSGW}$ and $\mathrm{QSG\hat{W}}$ cycles are iterated until the RMS change in the static part of quasiparticlized self-energy $\Sigma(0)$  reaches 10$^{-5}$ Ry. The two-particle Hamiltonian that is solved self-consistently to compute both the $\Sigma$ and the excitonic eigenvalues and eigenfunctions, contained 40 valence bands and 20 conduction bands. The necessity and sufficiency of such theories in describing both one- and two-particle transitions in a large class of 2D and 3D antiferromagnets in their ordered and disordered phases have been described in more detail in prior works~\cite{acharyaTheoryColorsStrongly2023,acharya2021electronic,acharya2022real,bianchi2023paramagnetic,watson2024giant}.

The exact diagonalization (ED)-dynamical mean field theory (DMFT) calculations were performed on top of a non-magnetic QSGW solution that contains two equivalent magnetic atoms in the unit cell. The five 3$d$ orbitals of the transition metal atom are included in the correlated Anderson impurity, while states within +10 and -10 eV around the Fermi energy are included in the bath. For Ni, Hubbard parameters are $U=4.2$~eV, $J=0.4$~eV, and a fully localized limit double counting correction is used, while for Mn $U=8$~eV and $J=0.8$~eV are used to get the band gaps consistent with the parameter-free estimations from of $\mathrm{QSG\hat{W}}$. Once the band gaps come out right, the higher order charge-charge correlators are computed exactly~\cite{acharyaTheoryColorsStrongly2023} from the ED solver. 

The ED-DMFT calculations are performed in single-site approximation. In this approach, using NiPS$_3$ as an example, Ni 3d  states constitute the correlated states. The projectors used for the Ni 3d orbitals are constructed by including all the bands from a window of $\pm$10~eV around the Fermi energy. By choosing a wide energy window, U becomes nearly static~\cite{choi2016first}. This window is used to build the hybridization matrix, which includes nickel, sulphur, and phosphorus states. However, hybridization is the kinetic energy part of the impurity Hamiltonian, while the U and J are explicitly included only for the Ni 3d states. Fluctuations of the one-particle states in the single-site DMFT approximations involve all processes that enable electrons and holes to move between the correlated Ni 3d states and the bath, which encompasses S, P, and other Ni orbitals. This is often referred to as $\mathrm{d^{n}\rightarrow d^{n\pm 1}}$ transition. However, this change in occupancy of the 3d states due to dynamical fluctuations is a strictly one-particle process. The excitonic (two-particle) processes in single-site approximation can only involve  $\mathrm{d^{n}\rightarrow d^{n}}$ transitions, where both the electron and the hole occupy the correlated Ni 3d site~\cite{acharyaTheoryColorsStrongly2023}. Consequently, the single-site DMFT cannot account for an excitonic state where the electron is on a correlated transition metal site and the hole is on the ligand (or vice-versa). These characteristics of the single-site DMFT approximation render the exciton to be purely onsite in character. In contrast, a Hamiltonian describing a NiS$_{6}$ cluster provides solutions, where the transition metal and the ligand share the electron and hole that can form a singlet exciton. Both single-site DMFT and the NiS$_{6}$ cluster calculations are constrained by approximations - exactly solvable single site DMFT does not incorporate the dipolar extended $dp$ excitonic term in the Hamiltonian while the cluster Hamiltonian for NiS$_{6}$ does not incorporate intersite $dd$ terms and also requires tuning of free parameters to adjust the exciton energies to their experimentally determined values. The most unbiased insight into the exciton comes from our $\mathrm{QSG\hat{W}}$ approach, which suggests that all transitions in this energy window (between 1 and 2 eV) contain onsite dipole forbidden $dd$, intersite dipole allowed $dd$, and $dp$ components. However, as the exciton energy increases (and it becomes weakly bound), the intersite components increase at the cost of the onsite component. However, in relative strengths, even for the 1.7 eV exciton (which is more weakly bound than the 1.47 eV exciton), all three processes contribute about 33\% each to the exciton wavefunction. This suggests that the observed excitons are neither strictly atom-local nor Zhang-Rice~\cite{PhysRevB.37.3759} states.

Within our QSGW+DMFT calculations for NiPS$_{3}$, we projected the lattice problem on the Ni d orbitals following the prescription of Haule~\cite{haule2010dynamical}. In order to single out the correlated subspace, a procedure of embedding, originally introduced in the above reference in the LAPW basis of the Wien2k package, is developed in the Full-Potential Linear Muﬃn-Tin Orbitals~\cite{methfessel2000full}. The technical information on the embedding process, choices of hybridization window, and the constrained RPA calculations performed to choose the Hubbard parameters are discussed in our previous work~\cite{PhysRevB.95.041112}.

\section{Data availability}
All data needed to evaluate the conclusions in this article are present herein and/or in the Supplementary Information Appendix.
   
\section{Acknowledgements}
 This project was supported by the Ministry of Education (Singapore) through the Research Centre of Excellence program (grant EDUN C-33-18-279-V12, I-FIM), and Academic Research Fund Tier 2 (MOE-T2EP50122-0012). This material is based upon work supported by the Air Force Office of Scientific Research and the Office of Naval Research Global under award number FA8655-21–1-7026. This work was authored in part by the National Renewable Energy Laboratory for the U.S. Department of Energy (DOE) under Contract No. DE-AC36-08GO28308. Funding was provided by the Computational Chemical Sciences program within the Office of Basic Energy Sciences, U.S. Department of Energy.  SA, DP and MvS acknowledge the use of the National Energy Research Scientific Computing Center, under Contract No. DE-AC02-05CH11231 using NERSC award BES-ERCAP0021783, and also acknowledge that a portion of the research was performed using computational resources sponsored by the Department of Energy's Office of Energy Efficiency and Renewable Energy and located at the National Renewable Energy Laboratory, and computational resources provided by the Oakridge leadership Computing Facility. The views expressed in the article do not necessarily represent the views of the DOE or the U.S. Government. The U.S. Government retains and the publisher, by accepting the article for publication, acknowledges that the U.S. Government retains a nonexclusive, paid-up, irrevocable, worldwide license to publish or reproduce the published form of this work, or allow others to do so, for U.S. Government purposes. M.P. acknowledges support from the CENTERA2, FENG.02.01-IP.05- T004/23 project funded within the IRA program of the FNP Poland, cofinanced by the EU FENG Programme
and from the ERC-AG TERAPLASM (No. 101053716) project.

\section{Author Contributions}
D.J. performed the data analysis, wrote the preliminary version of the manuscript, conducted the experiments, and conceptualized the work. S.A. performed the theoretical calculations, MvS and D.P provided the necessary software support, and S.A, MvS, and D.P contributed to the drafting of the paper. M.O., C.F., M.P., and M.K. discussed the results and contributed to setting the final version of the manuscript.

\bibliographystyle{ieeetr}
\providecommand{\noopsort}[1]{}\providecommand{\singleletter}[1]{#1}%

\newpage
\pagenumbering{gobble}

\begin{figure}[htp]
\includegraphics[page=1,trim = 18mm 18mm 18mm 18mm,
width=1.0\textwidth,height=1.0\textheight]{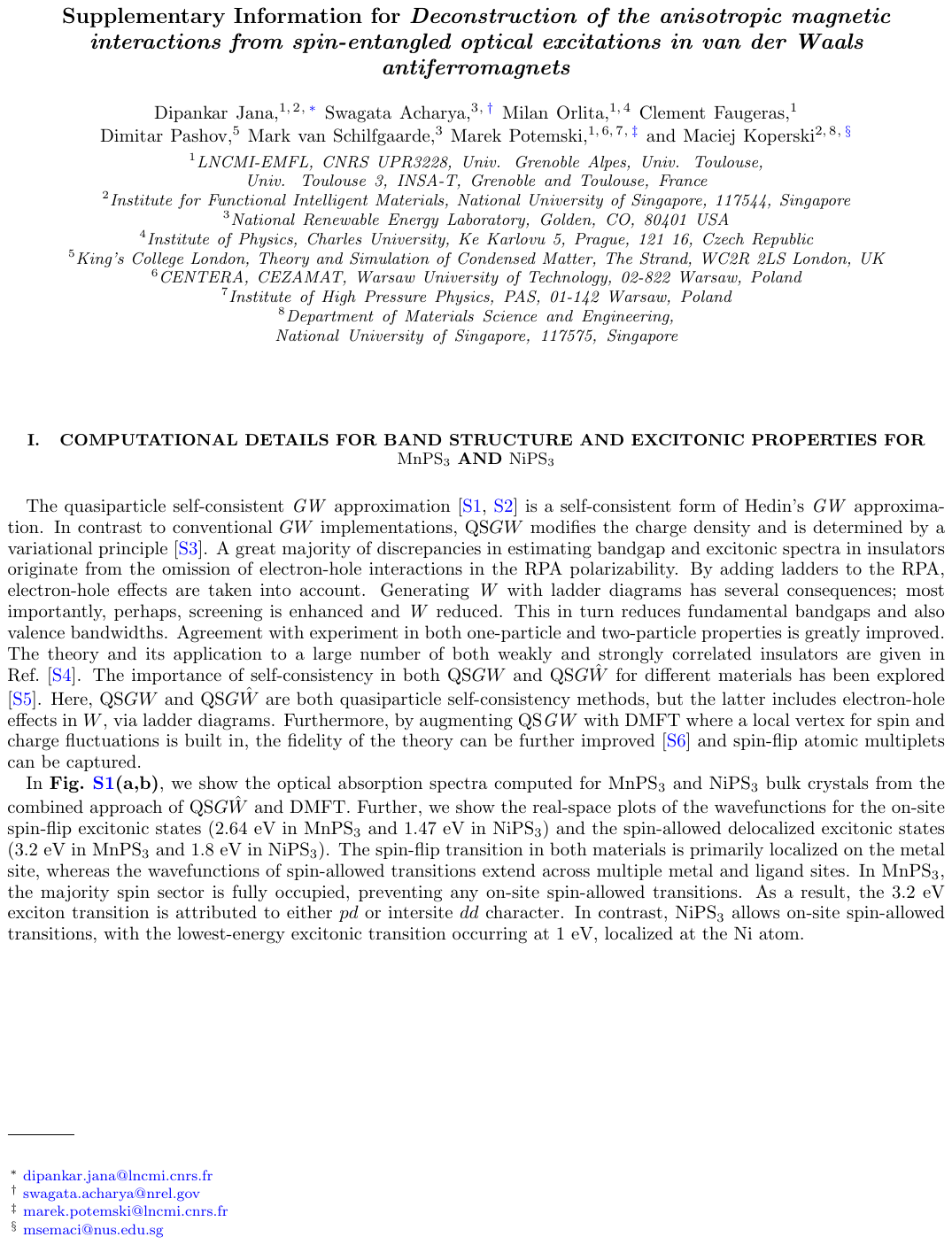}
\end{figure}

\newpage

\begin{figure}[htp]
   \includegraphics[page=2,trim = 18mm 18mm 18mm 18mm,
width=1.0\textwidth,height=1.0\textheight]{MnPS3_magnetic_phase_SM.pdf}
\end{figure}
\newpage

\begin{figure}[htp]
   \includegraphics[page=3,trim = 18mm 18mm 18mm 18mm,
width=1.0\textwidth,height=1.0\textheight]{MnPS3_magnetic_phase_SM.pdf}
\end{figure}

\begin{figure}[htp]
   \includegraphics[page=4,trim = 18mm 18mm 18mm 18mm,
width=1.0\textwidth,height=1.0\textheight]{MnPS3_magnetic_phase_SM.pdf}
\end{figure}

\newpage

\begin{figure}[htp]
   \includegraphics[page=5,trim = 18mm 18mm 18mm 18mm,
width=1.0\textwidth,height=1.0\textheight]{MnPS3_magnetic_phase_SM.pdf}
\end{figure}
\newpage

\begin{figure}[htp]
   \includegraphics[page=6,trim = 18mm 18mm 18mm 18mm,
width=1.0\textwidth,height=1.0\textheight]{MnPS3_magnetic_phase_SM.pdf}
\end{figure}

\begin{figure}[htp]
   \includegraphics[page=7,trim = 18mm 18mm 18mm 18mm,
width=1.0\textwidth,height=1.0\textheight]{MnPS3_magnetic_phase_SM.pdf}
\end{figure}

\begin{figure}[htp]
   \includegraphics[page=8,trim = 18mm 18mm 18mm 18mm,
width=1.0\textwidth,height=1.0\textheight]{MnPS3_magnetic_phase_SM.pdf}
\end{figure}

\end{document}